\documentclass[11pt,fleqn]{article}

\usepackage{amsmath,amssymb,amsthm,enumerate}

\newcommand{\p}{\partial}

\newcommand{\Equiv}{\mathop{ \sim}}

\newcommand{\sign}{\mathop{\rm sign}\nolimits}

\newcounter{tbn}

\newcounter{mcasenum}

\newtheorem{theorem}{Theorem}

{\theoremstyle{definition}

\newtheorem{note}{Note}
}

\begin{document}

\par\noindent {\LARGE\bf
Conservation laws and hierarchies of potential\\ symmetries for certain diffusion equations
\par}
{\vspace{4mm}\par\noindent {\bf N.M. Ivanova~$^\dag$, R.O. Popovych~$^\ddag$, C. Sophocleous~$^\diamondsuit$ and O.O. Vaneeva~$^\S$
} \par\vspace{2mm}\par}
{\vspace{2mm}\par\noindent {\it
$^\dag{}^\ddag{}^\S$~Institute of Mathematics of NAS of Ukraine, 
3 Tereshchenkivska Str., 01601 Kyiv, Ukraine\\
}}
{\noindent \vspace{2mm}{\it
$\phantom{^\dag{}^\ddag{}^\S}$~e-mail: ivanova@imath.kiev.ua, rop@imath.kiev.ua, vaneeva@imath.kiev.ua
}\par}

{\par\noindent\vspace{2mm} {\it
$^\ddag$~Fakult\"at f\"ur Mathematik, Universit\"at Wien, Nordbergstra{\ss}e 15, A-1090 Wien, Austria
} \par}

{\vspace{2mm}\par\noindent {\it
$^\diamondsuit$~Department of Mathematics and Statistics, University of Cyprus,
CY 1678 Nicosia, Cyprus\\
}}
{\noindent {\it
$\phantom{^\S}$~e-mail: christod@ucy.ac.cy
} \par}

{\vspace{5mm}\par\noindent\hspace*{8mm}\parbox{140mm}{\small
We show that the so-called hidden potential symmetries considered in a recent paper \cite{Gandarias2008}
are ordinary potential symmetries that can be obtained using the method
introduced by Bluman and collaborators \cite{Bluman&Kumei1989,Bluman&Reid&Kumei1988}.
In fact, these are simplest potential symmetries
associated with potential systems which are constructed with single conservation laws  having no constant characteristics.
Furthermore we classify the conservation laws for classes of porous medium equations and then using the corresponding
conserved (potential) systems we search for potential symmetries.
This is the approach one needs to adopt in order to determine the complete list of potential symmetries.
The provenance of potential symmetries is explained for the porous medium equations by using
potential equivalence transformations.
Point and potential equivalence transformations are also applied
to deriving new results on potential symmetries and corresponding invariant solutions from known ones.
In particular, in this way the potential systems, potential conservation laws and potential symmetries
of linearizable equations from the classes of differential equations
under consideration are exhaustively described.
Infinite series of infinite-dimensional algebras of potential symmetries are constructed for such equations.
}\par\vspace{5mm}}


\section{Introduction}

A Lie symmetry group of a system of differential equations is a group of transformations
that depend on continuous parameters and map any solution to another solution of the system.
While there is no existing general theory for solving nonlinear equations,
employment of the concept of Lie symmetry has been very helpful in determining new exact solutions.
Details on the theory of Lie symmetry groups and their applications to differential equations can be found in a number of textbooks.
See for example, \cite{Bluman&Kumei1989,Olver1993,Ovsiannikov1982}.

Bluman et al. \cite{Bluman&Kumei1989,Bluman&Reid&Kumei1988} introduced a method for finding a new class of symmetries for
a system of partial differential equations $\Delta(x, u)$, in the case that this system has at least one conservation law.
If we introduce potential variables $v$ for the equations written
in conserved forms as further unknown functions, we obtain a system $Z(x,u,v)$.
Any Lie symmetry for $Z(x,u,v)$ induces a  symmetry for $\Delta(x, u)$.
When at least one of the infinitesimals which correspond to the variables $x$ and $u$ depends explicitly on
potentials, then the local symmetry of $Z(x,u,v)$ induces a nonlocal symmetry of
$\Delta(x,u)$. These nonlocal symmetries are called {\it potential symmetries}. More details about
potential symmetries and their applications can be found in \cite{Bluman&Kumei1989,bluman93a,bluman93b}.
Potential symmetries were investigated for quite general classes of differential equations.
The problem of finding criteria for the existence of potential symmetries
for classes of differential equations was posed in~\cite{Pucci&Saccomandi1993}.
Some useful criteria were derived for partial differential equations in two independent variables.
Nonclassical potential symmetries of such equations were discussed in~\cite{Saccomandi1997}.

For partial differential equations in two independent variables, $t$ and $x$, the general form of (local) conservation laws is
\begin{equation}\label{conslaw}
D_tF+D_xG=0,
\end{equation}
where $D_t$ and $D_x$ are the total derivatives with respect to $t$ and $x$.
The equality~\eqref{conslaw} is assumed to be satisfied for any solution of the corresponding system of equations.
The components $F$ and $G$ of the conserved vector $(F,G)$ are functions of~$t$, $x$ and derivatives of~$u$ and
are called the {\it density} and the {\it flux} of the conservation law.
Basic definitions and statements on conservation laws
(conserved vector, equivalence of conserved vectors, characteristic, Noether's theorem etc.)
are presented, e.g., in~\cite{Olver1993}.

In the recent paper \cite{Gandarias2008} the construction of hidden potential symmetries for some classes of diffusion equations is claimed.
Here we show that these symmetries are usual potential symmetries that can be derived
using the conventional method by Bluman and collaborators.
In fact, the way proposed in \cite{Gandarias2008} for obtaining potential systems is a particular case of finding characteristics
of conservation laws of order zero and the so-called hidden potential symmetries are ordinary simplest potential symmetries
corresponding to potential systems constructed with conservation laws whose characteristics are nonconstant.
These symmetries are simplest since each of them is associated with a single conservation law
and hence involves only a single potential.
Furthermore potential symmetries derived in \cite{Gandarias2008} were also found in \cite{Sophocleous2005}.
Essentially more general results on potential conservation laws and potential symmetries of wider classes of porous medium equations
including that with an arbitrary number of potentials were obtained in
\cite{Ivanova&Popovych&Sophocleous2006Part3,Ivanova&Popovych&Sophocleous2006Part4,Popovych&Kunzinger&Ivanova2008}.
A procedure of classification of related quasilocal symmetries was proposed in~\cite{Zhdanov&Lahno2005}
for the general class of (1+1)-dimensional evolution equations.

In the subsequent analysis, we examine each diffusion equation considered in \cite{Gandarias2008} in more detail.
We also study a class of porous medium equations which was stated in \cite{Gandarias2008}.
Special potential symmetries for a partial case of such equations were earlier derived in \cite{Gandarias1996}.

The first step in the investigation of potential symmetries is to calculate the conservation laws.
The conventional symmetry approach for this is based on Noether's theorem but it cannot be directly used
for evolution equations.
There exists no Lagrangian for which an evolution equation is an Euler--Lagrange equation.
Hence the application of Noether's theorem in this case is possible only for particular equations and after
special technical tricks.
At~the same time, the definition of conservation laws itself gives rise to a method of finding conservation laws,
which is called \emph{direct} and can be applied to any system of differential equations with no restriction on its structure.
The technique of calculations used within the framework of this method is similar to
the classical Lie method yielding symmetries of differential equations.
Four versions of it are distinguished in the literature
depending on the way of taking into account systems under consideration
and the usage either the definition of conserved vectors or the characteristic form of conservation laws.
See, e.g., \cite{Anco&Bluman2002a,Anco&Bluman2002b,Popovych&Ivanova2004ConsLawsLanl} on details of the calculation technique
as well as~\cite{Wolf2002} for comparison of the versions and their realizations in computer algebra programs.
The necessary theoretical background is given in~\cite{Olver1993}.
In the present work we employ the most direct version \cite{Popovych&Ivanova2004ConsLawsLanl}
based on immediate solving of determining equations for
conserved vectors of conservations laws on the solution manifolds of investigated systems
and additionally combined with techniques involving symmetry or equivalence transformations.

A complete classification of potential symmetries can be achieved by considering
all potential systems that correspond to the conservation laws.
It is known~\cite{Popovych&Ivanova2004ConsLawsLanl} that the equivalence group for a class of systems of differential equations
or the symmetry group for a single system can be prolonged to potential variables. It is natural to use these prolonged
equivalence groups for classification of possible potential symmetries.
In view of this statement we will classify potential symmetries of diffusion equations
up to the (trivial) prolongation of their equivalence groups to the corresponding potentials.

In the next four sections we use classes of porous medium equations as examples and we find the first generation of potential symmetries.
In the end of section~\ref{SectionDifEq} and in section~\ref{SectionOnApplicationsOfPETs}
we present applications of potential equivalence transformations. In particular, a connection between
potential and Lie symmetries is established, new exact solutions are found
and exhaustive description of potential symmetries of linearizable diffusion equations is given.
Infinite series of infinite-dimensional algebras of potential symmetries are constructed for such equations.

\section{Fokker--Planck equations}\label{SectionFokPlEq}

The class of the Fokker--Planck equations
\begin{equation}\label{eqFokPlEq}
u_t=u_{xx}+[f(x)u]_x
\end{equation}
is contained in the class of (1+1)-dimensional second-order linear evolution equations
which have the general form
\begin{equation}\label{EqGenLPE}
u_t=A(t,x)u_{xx}+B(t,x)u_x+C(t,x)u,
\end{equation}
where the coefficients $A$, $B$ and $C$ are smooth functions of~$t$ and~$x$, $A\ne 0$.
Due to its importance and relative simplicity, class~\eqref{EqGenLPE} is
the most investigated in the framework of group analysis of differential equations.
Equations from this class are often used
as examples for the first presentations of investigations on different new kinds of symmetries,
illustrative examples in textbooks on the subject
and benchmark examples for computer programs calculating symmetries of differential equations.

\looseness=-1
In fact, the complete group classification of (1+1)-dimensional linear parabolic equations
(i.e., the complete description of their Lie symmetries up to the equivalence relation
generated by the corresponding equivalence group) was performed by Sophus Lie~\cite{Lie1881} himself
as a part of the more general group classification of
linear second-order partial differential equations in two independent variables.
A modern treatment of the subject is given in~\cite{Ovsiannikov1982}.
There exist also a number of recent papers partially rediscovering
the classical results of Lie and Ovsiannikov.
The local conservation laws of equations from class~\eqref{EqGenLPE} were calculated in~\cite{Bryant&Griffiths1995b}
using differential forms.
This result was reobtained in~\cite{Popovych&Kunzinger&Ivanova2008} by the direct method of calculations of conservation laws.

First the whole potential symmetry frame over class~\eqref{EqGenLPE}
including local and potential conservation laws and usual and generalized potential symmetries
with an arbitrary number of independent potentials of any level
was investigated in~\cite{Popovych&Kunzinger&Ivanova2008}
with combining a number of sophisticated techniques such as
a rather intricate interplay between different representations of potential systems,
the notion of a potential equation associated with a tuple of characteristics,
prolongation of the equivalence group to the whole potential frame and
application of multiple dual Darboux transformations.
In particular, all possible potential conservation laws of equations from
class~\eqref{EqGenLPE} proved to in fact be exhausted by local conservation laws.
Based on the tools developed, a preliminary analysis of generalized potential symmetries
is carried out and then applied to substantiate the choice of canonical forms for potential systems.
Effective criteria for the existence of potential symmetries are proposed for the case
of an arbitrary number of involved potentials.
Equations from class~\eqref{EqGenLPE}, possessing infinite series of potential symmetry algebras, are studied in detail.

Nonclassical (conditional) symmetries of equations from class~\eqref{EqGenLPE}
and all the possible reductions of these equations to ordinary differential ones
are exhaustively described in~\cite{Popovych2008a}.
This problem proves to be equivalent, in some sense, to solving the initial equations.
The ``no-go'' result is extended to the investigation of point transformations
(admissible transformations, equivalence transformations, Lie symmetries) and Lie reductions
of the determining equations for the nonclassical symmetries.

There exist a number of earlier papers on investigation of particular potential or nonclassical
symmetries of narrow subclasses of equations from class~\eqref{EqGenLPE} or even single equations
from this class (see references in~\cite{Popovych2008a,Popovych&Kunzinger&Ivanova2008}).
Below we discuss only some results which are directly related to class~\eqref{eqFokPlEq}.
Note that any pair $(\mathcal L,\mathcal F)$, where $\mathcal L$ is an equation from class~\eqref{EqGenLPE} and
$\mathcal F$ is a conservation law of it, is reduced by a point transformation from the equivalence group of
class~\eqref{EqGenLPE} to a pair $(\tilde{\mathcal L},\tilde{\mathcal F})$,
where $\tilde{\mathcal L}$ is a Fokker--Planck equation $u_t=u_{xx}+(F(t,x)u)_x$ and
$\tilde{\mathcal F}$ is its conservation law with the characteristic~1 
(i.e., with the characteristic which is identically equal to~1) \cite[Proposition~8]{Popovych&Kunzinger&Ivanova2008}.

Class~\eqref{eqFokPlEq} admits the generalized extended equivalence group formed by the transformations
\begin{gather*}\tilde t=\delta_1^2t+\delta_2,\quad\tilde x=\delta_1x+\delta_3,
\quad\tilde
u=\psi u+\psi\zeta,\quad
\tilde f=
\frac f{\delta_1}-\frac2{\delta_1}\frac{\psi_{x}}{\psi},\end{gather*}
where $\psi=\delta_4\int \!{e^{\int \!f(x)dx}}dx+\delta_5;$
$\delta_i$, $i=1,\dots,5,$ are arbitrary constants, $\delta_1(\delta_4^2+\delta_5^2)\ne0;$
$\zeta=\zeta(t,x)$ is an arbitrary solution of~\eqref{eqFokPlEq}.
Each local conservation law of~\eqref{eqFokPlEq} is canonically represented by a conserved vector $(\alpha u,-\alpha u_x+(\alpha_x-\alpha f)u)$,
where $\alpha=\alpha(t,x)$ is an arbitrary solution of the adjoint equation $\alpha_t+\alpha_{xx}-f\alpha_x=0$ \cite{Pucci&Saccomandi1993b}.
Then, in view of Theorem~5 and Corollaries~20 and~30 of~\cite{Popovych&Kunzinger&Ivanova2008}
any potential system of~\eqref{eqFokPlEq}, which is essential for finding potential symmetries,
has the~form
\begin{equation}\label{EqPotSystemsForFokPlEq}
v^i_x=\alpha^iu,\quad v^i_t=\alpha^i u_x-(\alpha^i_x-\alpha^if)u, \quad i=1,\ldots,k,
\end{equation}
where $k\in\mathbb N$
and $\alpha^i=\alpha^i(t,x)$, $i=1,\ldots,k$, are arbitrary linearly independent solutions of the adjoint equation.
The same statement for the particular case of the linear heat equation was earlier obtained in~\cite{Popovych&Ivanova2004ConsLawsLanl}.

In~\cite{Pucci&Saccomandi1993b} the potential systems of the form~\eqref{EqPotSystemsForFokPlEq}
with a single potential (the case $k=1$) were considered.
Their Lie symmetries were preliminary investigated.
The so-called natural potential systems associated with the characteristic~1 were studied in more detail.
As a result, the corresponding special simplest potential symmetries were completely classified.
(We call \emph{simplest potential symmetries} ones arising under the consideration of potential system with a single potential.)
The potential symmetries of the same kind were also found in~\cite{Pucci&Saccomandi1993}
for the particular equation of form~\eqref{eqFokPlEq} with $f=x$, i.e., $u_t=u_{xx}+(xu)_x$.
It was shown in~\cite{Ivanova&Popovych2007CommentOnMei} that these potential symmetries
are obtained from the ``natural'' simplest potential symmetries of the linear heat equation $u_t=u_{xx}$
via a point transformation connecting the equations $u_t=u_{xx}$ and $u_t=u_{xx}+(xu)_x$.

Recently these results were essentially generalized in~\cite{Popovych&Kunzinger&Ivanova2008}.
Any characteristic of the linear heat equations, which gives rise to nontrivial simplest potential symmetries,
proves to be equivalent, with respect to the corresponding Lie symmetry group,
to either the characteristic~1 or the characteristic~$x$~\cite[Theorem~7]{Popovych&Kunzinger&Ivanova2008}.
The Lie algebra of simplest potential symmetries of the linear heat equation, connected with the characteristic~$x$,
was calculated in~\cite{Ivanova&Popovych2007CommentOnMei}.
The above point transformation allows us to easily extend the description of characteristics which are essential
for finding simplest potential symmetries to the Fokker-Planck equation $u_t=u_{xx}+(xu)_x$.
Namely, any nontrivial simplest potential symmetry of the equation  $u_t=u_{xx}+(xu)_x$
is associated, up to equivalence generated by the corresponding Lie symmetry group,
with either the characteristic~1 or the characteristic~$e^tx$.
The Lie symmetry algebras of the potential systems constructed with the above characteristics are calculated.
In other words, in~\cite{Popovych&Kunzinger&Ivanova2008} all simplest potential symmetries
of the linear heat equation $u_t=u_{xx}$ and the Fokker-Planck equation $u_t=u_{xx}+(xu)_x$ were found.
Moreover, these equations are shown to possess infinite series of algebras of potential symmetries
depending on an arbitrary number of potentials~\cite[Proposition~13]{Popovych&Kunzinger&Ivanova2008}.
Other equations from class~\eqref{EqGenLPE} (including those from class~\eqref{eqFokPlEq}) possessing such series are singled out.

System~(43) of~\cite{Gandarias2008} is a partial case of the potential system~\eqref{EqPotSystemsForFokPlEq}
with $k=1$ and $\alpha$ depending only on~$x$.
As above shown, such systems were intensively investigated earlier and/or in a more general frame than in~\cite{Gandarias2008}.

Results discussed in this section are easily extended by point or nonlocal equivalence transformations
to linearizable second-order evolution equations~\cite{Popovych&Ivanova2004ConsLawsLanl}.
Examples of such extension are presented in Sections~\ref{SectionDifEq} and~\ref{SectionOnApplicationsOfPETs}.
All the possible potential conservation laws  of the corresponding linearizable equations
and their potential symmetries depending on an arbitrary number of potentials are exhaustively described in this way.

\section{Inhomogeneous diffusion equations}\label{SectionDifEq}

In this section we consider variable coefficient nonlinear diffusion equations of the general form
\begin{equation}\label{eqDifEqfg}
f(x)u_t=(g(x)u^nu_x)_x.
\end{equation}
Using the transformation $\tilde t=t$, $\tilde x=\int \frac{dx}{g(x)}$, $\tilde u=u$,
we can reduce equation~\eqref{eqDifEqfg} to
\begin{equation*}
\tilde f(\tilde x)\tilde u_{\tilde t}= (\tilde u^n \tilde u_{\tilde x})_{\tilde x},
\end{equation*}
where $\tilde f(\tilde x)=g(x)f(x)$ and $\tilde g(\tilde x)\equiv1$.
That is why, without loss of generality, we restrict ourselves to the investigation of equations
having the form
\begin{equation}\label{eqDifEqf}
f(x)u_t=(u^nu_x)_x.
\end{equation}
The gauge $g=1$ could be replaced by other gauges of arbitrary elements.
For example, any equation of form~\eqref{eqDifEqfg} can be reduced
by a transformation similar to the above one to an equation of the same form with~$f=1$.
The gauge~$f=1$ is conventionally used for arbitrary elements of class~\eqref{eqDifEqfg}
but the application of the gauge $g=1$ instead of $f=1$ leads to simplifying all investigations of
symmetry and transformational properties of class~\eqref{eqDifEqfg} and hence is optimal.
The usage of different gauges is discussed in~\cite{Ivanova&Popovych&Sophocleous2006Part1,VJPS2007,VPS2008}.

The equivalence group~$G^{\sim}$ of class~\eqref{eqDifEqf} has a simple structure and consists of the transformations
\begin{equation} \label{EquivTransformationsDifEqfg}\arraycolsep=0em
\tilde t=\delta_1 t+\delta_4,\quad
\tilde x=\delta_2 x+\delta_5, \quad
\tilde u=\delta_3 u,\quad
\tilde f=\delta_1\delta_2^{-2}\delta_3^nf, \quad
\tilde n=n,
\end{equation}
where $\delta_i$, $i=1,\dots,5$, are arbitrary constants, $\delta_1\delta_2\delta_3\not=0$.
At the same time, class~\eqref{eqDifEqf} possesses a generalized equivalence group
which is wider than~$G^{\sim}$.
The notion of generalized equivalence groups was proposed by Meleshko~\cite{Meleshko1994}.
See also~\cite{Ivanova&Popovych&Sophocleous2006Part1,Popovych&Kunzinger&Eshraghi2006,VPS2008}
for discussions and generalizations of this notion as well as a number of examples of classes
having nontrivial generalized equivalence groups and ways to use them in solving different classification problems.
In contrast to usual equivalence groups~\cite{Ovsiannikov1982},
components of transformations from generalized equivalence groups,
associated with independent variables and unknown functions, may depends on arbitrary elements of the corresponding classes.

\begin{theorem}
The generalized equivalence group~$\hat G^{\sim}$ of
class~\eqref{eqDifEqf} under the condition $n\ne-1$ consists of the transformations
\[\hspace*{-.5\arraycolsep}
\begin{array}{l}
\tilde t=\delta_1 t+\delta_2,\quad \tilde x=\dfrac{\delta_3x+\delta_4}{\delta_5x+\delta_6}, \quad
\tilde u=\delta_7|\delta_5x+\delta_6|^{-\frac1{n+1}}u, \\[2ex]
\tilde f=\delta_1{\delta_7}^n|\delta_5x+\delta_6|^{\frac{3n+4}{n+1}} f, \quad
\tilde n=n.
\end{array}
\]
where $\delta_j$, $j=1,\dots,7$, are arbitrary constants,
$\delta_1\delta_7\not=0$ and $\delta_3\delta_6-\delta_4\delta_5=\pm1$.

In the case $n=-1$ transformations from the group~$\hat G^{\sim}$ take the form
\[
\tilde t=\delta_1 t+\delta_2,\quad
\tilde x=\delta_3 x+\delta_4,\quad
\tilde u=\delta_5e^{\delta_6 x} u,\quad
\tilde f=\delta_1\delta_3^{-2}\delta_5^{-1}e^{-\delta_6 x}f,
\]
where $\delta_j$, $j=1,\ldots,6$, are arbitrary constants, $\delta_1\delta_3\delta_5\not=0$.
\end{theorem}

Since the parameter~$n$ is an invariant of all admissible (point) transformations in class~\eqref{eqDifEqf},
this class can be presented as the union of disjoint subclasses where
each from the subclasses corresponds to a fixed value of~$n$.
This representation allows us to give the interpretation of the generalized equivalence group~$\hat G^{\sim}$
as a family of the usual conditional equivalence groups of the subclasses parameterized with~$n$,
and the value $n=0$ and $n=-1$ being singular.
In the case $n=0$ a part of the corresponding conditional equivalence group has to be neglected.

The conservation laws for the class~\eqref{eqDifEqf} are stated in the following theorem
\cite{Ivanova&Popovych&Sophocleous2004,Ivanova&Popovych&Sophocleous2006Part3}.

\begin{theorem}
The space of local conservation laws of any equation of form~\eqref{eqDifEqf} with~$n\ne0$ is two-dimensional and spanned by conservation laws with
the conserved vectors~$(fu, -u^nu_x)$ and $(\,xfu,\ -xu^nu_x+\int\!\! u^ndu\,)$.
The space of local conservation laws of the linear equation $fu_t=u_{xx}$ ($n=0$) is infinite-dimensional
and spanned by $(\,\alpha f u, \ -\alpha u_x+\alpha_x u\,)$.
Here $\alpha=\alpha(t,x)$ runs through the solution set of the linear equation \mbox{$f\alpha_t+\alpha_{xx}=0$}.
\end{theorem}

Up to $G^\sim$-equivalence,
these conservation laws give rise to the following inequivalent potential systems for equations~\eqref{eqDifEqf} with~$n\ne0$:

\bigskip

\noindent{\bf 1.}  \quad $v^1_x=fu$,\quad $v^1_t=u^nu_x$;
\\[1ex]
\noindent{\bf 2.} \quad $v^2_x=xfu$,\quad $v^2_t=xu^nu_x-\int\! u^ndu$;
\\[1ex]
\noindent{\bf 3.} \quad  $v^1_x=fu$,\quad $v^1_t=u^nu_x$,\quad $v^2_x=xfu$,\quad $v^2_t=xu^nu_x-\int\! u^ndu$.

\bigskip

\noindent
Systems~{\bf1} and~{\bf2} are associated with the conservation laws having the characteristics~1 and~$x$, respectively.
The united system~{\bf3} corresponds to the whole space of conservation laws.
The generalized equivalence group~$\hat G^{\sim}$ prolonged to potentials establishes additional equivalence between potential systems.
Thus, in the case $n\ne-1$ the transformation
\begin{equation}\label{EqGenEquivTransForEqDifEqf}
\tilde t=t, \quad
\tilde x=x^{-1}, \quad
\tilde u=|x|^{-\frac1{n+1}}u, \quad
\tilde v^1=-(\sign x)v^2, \quad
\tilde v^2=-(\sign x)v^1
\end{equation}
maps systems~{\bf1} and~{\bf2} to systems~{\bf2} and~{\bf1} in the tilde variables
with $\tilde f=|\tilde x|^{-\frac{3n+4}{n+1}}f(\tilde x^{-1})$, respectively.
Systems~{\bf1} and~{\bf2} are $\hat G^\sim$-inequivalent for an arbitrary pair of values of~$f$ iff $n=-1$.

System~(17) of~\cite{Gandarias2008} with $g=x^{\pm2}$ is a particular case of the above system {\bf2},
and all symmetries obtained in~\cite{Gandarias2008} are nothing but usual potential symmetries
derived previously in~\cite{Sophocleous2005}.
Thus, the operators $\boldsymbol{{\rm v}_4}$ and $\boldsymbol{{\rm v}_6}$ from~(20) in~\cite{Gandarias2008} coincide with
the operators~$\Gamma_2/2$ and~$\Gamma_1/4$, where the operators~$\Gamma_2$ and~$\Gamma_1$
are presented in formulas~(5.7) and~(5.6) of~\cite{Sophocleous2005}, respectively.
The operator $\boldsymbol{{\rm w}_4}$ given in~(25) of~\cite{Gandarias2008} differs from the operator~$\Gamma_1/3$ by scaling
the corresponding potential, where the operator~$\Gamma_1$ is defined in~(5.9) of~\cite{Sophocleous2005}.
Additionally note that system~(26) of~\cite{Gandarias2008} does not define the potential in a proper way.

Potential symmetries of equation~\eqref{eqDifEqf}, associated with system~{\bf1}, were first obtained in~\cite{Sophocleous2000,Sophocleous2003},
see also~\cite{Akhatov&Gazizov&Ibragimov1989,Bluman&Kumei1989,Bluman&Reid&Kumei1988} for the constant coefficient case~$f=1$.
There exist two inequivalent equations of form~\eqref{eqDifEqf} admitting such nonlocal symmetries.
Below we adduce the values of arbitrary elements together with bases of the corresponding maximal Lie invariance algebras.

\bigskip

\noindent1.1. $f=1$, $n=-2$:\\[1ex]
\phantom{1.1.} $\langle \p_t,\ $ $\p_{v^1},\ $ $2t\p_t+u\p_u+v^1\p_{v^1},\  x\p_x-u\p_u,\  -v^1x\p_x+(xu+v^1)u\p_u+2t\p_{v^1},$\\[1ex]
\phantom{1.1.} $4t^2\p_t-((v^1)^2+2t)x\p_x+((v^1)^2+6t+2xuv^1)u\p_u+4tv^1\p_{v^1},\  \varphi\p_x-\varphi_{v^1}u^2\p_u\rangle$;

\bigskip

\noindent1.2. $f=x^{-4/3}$, $n=-2$:\\[1ex]
\phantom{1.2.} $\langle\p_t,\ \p_{v^1},\  2t\p_t+u\p_u+v^1\p_{v^1},\
3x\p_x-u\p_u-2v^1\p_{v^1},\  3xv^1\p_x-(v^1+3x^{-1/3}u)u\p_u-(v^1)^2\p_{v^1}\rangle.$

\bigskip

\noindent
Here and below $\varphi=\varphi(t,v^1)$ is an arbitrary solution of the linear heat equation $\varphi_t=\varphi_{v^1v^1}$.

Potential symmetries of equation~\eqref{eqDifEqf} associated with system~{\bf2} were first investigated in~\cite{Sophocleous2005}
(see also~\cite{Ivanova&Popovych&Sophocleous2006Part4}).
Up to the equivalence group~$G^{\sim}$, there exist exactly two cases of equations in class~\eqref{eqDifEqf} admitting such potential symmetries:

\bigskip

\noindent2.1. $f=x^{-2}$, $n=-2$:\\[1ex]
\phantom{2.1.} $\langle \p_t,\ \p_{v^2},\ x\p_x,\ 2t\p_t+u\p_u+v^2\p_{v^2},\ v^2x\p_x-u^2\p_u+2t\p_{v^2},$\\[1ex]
\phantom{2.1.} $4t^2\p_t+((v^2)^2+2t)x\p_x+2(2t-uv^2)u\p_u+4tv^2\p_{v^2},\ x^2\psi\p_x-xu(\psi+\psi_{v^2}u)\p_u\rangle$.

\bigskip

\noindent2.2. $f=x^{-2}(c_1+c_2x^{-1})^{-4/3}$,\quad $c_2\neq0$,\quad  $n=-2$:\\[1ex]
\phantom{2.2.} $\langle\p_t,\ \p_{v^2},\ 2t\p_t+u\p_u+v^2\p_{v^2},\  3(c_1x+c_2)x\p_x-(2c_2+3c_1x)u\p_u+2c_2v^2\p_{v^2},$\\[1ex]
\phantom{2.2.} $3v^2(c_1x+c_2)x\p_x-(3x^{4/3}(c_1x+c_2)^{-1/3}u+(2c_2+3c_1x)v^2)u\p_u+c_2(v^2)^2\p_{v^2} \rangle$;

\bigskip

\noindent
Here and below $\psi=\psi(t,v^2)$ is an arbitrary solution of the linear heat equation $\psi_t=\psi_{v^2v^2}$.
By transformation~\eqref{EqGenEquivTransForEqDifEqf}, cases~2.1 and~2.2 are reduced to cases~1.1 and~1.2, respectively.
For the precise reduction $2.2\to1.2$ the transformation $\hat t=\tilde t$, $\hat x=c_1+c_2\tilde x$, $\hat u=c_2^{-1}\tilde u$
from $G^\sim$ has to be additionally carried out.

The united system~{\bf3} is equivalent to the second-level potential system
\begin{equation}\label{SystemPotSysSecondLevel}
\textstyle
v^1_x=fu,\quad w_x=v^1,\quad w_t=\int\! u^ndu  
\end{equation}
constructed from system~{\bf1} using its conserved vector $(v,-\int\! u^ndu)$, and $w=xv^1-v^2$.
Nontrivial $G^\sim$-inequivalent cases of potential symmetries associated with system~\eqref{SystemPotSysSecondLevel}
are exhausted by the following ones:

\bigskip

\noindent3.1. $f=1$, $n={-2}$:\\[1ex]
\phantom{3.1.} $\langle\p_t,\  \p_w,\  \p_{v^1}+x\p_w,\ 2t\p_t+u\p_u+v^1\p_{v^1}+w\p_w,\  x\p_x-u\p_u+w\p_w$,\\[1ex]
\phantom{3.1.} $(w-2v^1x)\p_x+(2xu+v^1)u\p_u+2t\p_{v^1}+(2t-(v^1)^2)x\p_w$,\\[1ex]
\phantom{3.1.} $4t^2\p_t+(2v^1w-3x(v^1)^2-6tx)\p_x+(6xuv^1-2uw+(v^1)^2+10t)u\p_u+4tv^1\p_{v^1}$\\[1ex]
\phantom{3.1.} ${}+((v^1)^2w-2tw-2x(v^1)^3)\p_w,\ \varphi_{v^1}\p_x-\varphi_tu^2\p_u+(v^1\varphi_{v^1}-\varphi)\p_w\rangle$;

\bigskip

\noindent3.2. $f=1$, $n={-2/3}$:\\[1ex]
\phantom{3.2.} $\langle\p_t,\  \p_x,\   \p_w,\  \p_{v^1}+x\p_w,\  2t\p_t+3u\p_u+3v^1\p_{v^1}+3w\p_w,\
x\p_x-3u\p_u-2v^1\p_{v^1}-w\p_w,\ $\\[1ex]
\phantom{3.2.} $w\p_x-3uv^1\p_u-(v^1)^2\p_{v^1}\rangle$;

\bigskip

\noindent3.3. $f=x^{-2}$, $n={-2}$: \\[1ex]
\phantom{3.3.} $\langle\p_t,\ \p_w,\ \p_{v^1}+x\p_w,\ x\p_x-v^1\p_{v^1},\ 2t\p_t+u\p_u+v^1\p_{v^1}+w\p_w,$\\[1ex]
\phantom{3.3.} $x(2xv^1-w)\p_x-u(xv^1+2u)\p_u+v^1(w-xv^1)\p_{v^1}+(x^2(v^1)^2-2t)\p_w,$\\[1ex]
\phantom{3.3.} $4t^2\p_t+x(6t+3x^2(v^1)^2-4xv^1w+w^2)\p_x+2u(2t-3xuv^1+2uw-x^2(v^1)^2+xv^1w)\p_u$\\[1ex]
\phantom{3.3.} $+(2xv^1w-2t-x^2(v^1)^2-w^2)v^1\p_{v^1}+2(2tw+x^3(v^1)^3-x^2(v^1)^2w)\p_w,$\\[1ex]
\phantom{3.3.} $x^2\psi_{v^2}\p_x+(\psi_{v^2}+u\psi_t)xu\p_u-\psi\p_{v^1}+x(xv^1\psi_{v^2}-\psi)\p_w\rangle$;

\bigskip

\noindent3.4. $f=x^{-6}$, $n={-2/3}$: \\[1ex]
\phantom{3.4.} $\langle\p_t,\ \p_{w},\ \p_{v^1}+x\p_w,\ 2t\p_t+3u\p_u+3v^1\p_{v^1}+3w\p_w,\
x\p_x+6u\p_u+v^1\p_{v^1}+2w\p_w,$\\[1ex]
\phantom{3.4.} $x^2\p_x+3xu\p_u+(w-xv^1)\p_{v^1}+xw\p_w,\ xw\p_x-3(xv^1-2w)u\p_u-(xv^1-w)v^1\p_{v^1}+w^2\p_w\rangle$.

\bigskip

Here $v^2=xv^1-w$.
By transformation~\eqref{EqGenEquivTransForEqDifEqf} which is rewritten for~$v^1$ and~$w$ as $\tilde v^1=w\sign x-|x|v^1$ and $\tilde w=|x|^{-1}w$,
cases~3.3 and~3.4 are reduced to cases~3.1 and~3.2, respectively.

The linear equation $fu_t=u_{xx}$ ($n=0$) possesses an infinite series of potential systems of the form

\bigskip

\noindent
{\bf 4.} \quad $v^i_x=\alpha^ifu$,\quad $v^i_t=\alpha^i u_x-\alpha^i_x u$,\quad $i=1,\dots,k$,

\bigskip

where $k\ge1$ and $\alpha^i=\alpha^i(t,x)$ are arbitrary linearly independent solutions of the linear equation \mbox{$f\alpha_t+\alpha_{xx}=0$}.
For the classification of potential symmetries of linear parabolic equations
(in particular, the investigation of Lie symmetries of system~{\bf4}) we refer the reader to~\cite{Popovych&Kunzinger&Ivanova2008}.
See also the previous section for the discussion of this result.

The appearance of nontrivial potential symmetries for equations from class~\eqref{eqDifEqf}
can be easily explained using transformations involving potentials and called potential equivalence transformations (PETs),
cf.~\cite{Popovych&Ivanova2005PETs}.
It is sufficient to consider only the potential hodograph transformation
\begin{equation}\label{EqHodographXV}
\tilde t=t, \quad \tilde x=v^1, \quad \tilde u=u^{-1}, \quad \tilde v^1=x,
\end{equation}
and $\tilde v^2=v^2$ (resp. $\tilde w=w$).
Under transformation~\eqref{EqHodographXV}, system~{\bf1} is mapped to the system
\begin{equation}\label{EqTransformedPotSysForEqDifEqf}
f(v^1)\tilde v^1_{\tilde x}=\tilde u, \quad \tilde v^1_{\tilde t}=\tilde u^{-n-2}\tilde u_{\tilde x}.
\end{equation}
Since transformation~\eqref{EqHodographXV} is a point transformation in the space of variables supplemented with the potentials,
it establishes an isomorphism between maximal Lie invariance algebras of systems~{\bf1} and~\eqref{EqTransformedPotSysForEqDifEqf}.
Note that transformation~\eqref{EqHodographXV} is an involution, i.e., it coincides with its inverse.

If $f=1$ and $n=-2$, system~\eqref{EqTransformedPotSysForEqDifEqf} coincides with the potential system
of the linear heat equation $\tilde u_{\tilde t}=\tilde u_{\tilde x\tilde x}$, associated with the characteristic~1.
The potential systems of the linear heat equation are exhausted by the potential systems of form~{\bf4} with~$f=1$
\cite{Popovych&Ivanova2004ConsLawsLanl,Popovych&Kunzinger&Ivanova2008}.
Therefore, the whole set of potential systems  (of all levels) of the $u^{-2}$-diffusion equation $u_t=(u^{-2}u_x)_x$ consists of
the images of the potential systems of the linear heat equation, constructed with the tuples of characteristics
including the characteristic~1 as the first element, with respect to transformation~\eqref{EqHodographXV}
identically prolonged to the other potentials.
The maximal level of potentials and potential systems equals two.
All properties of potential symmetries of the $u^{-2}$-diffusion equation follow from the corresponding properties
of the linear heat equation.
In particular, Proposition 12 of \cite{Popovych&Kunzinger&Ivanova2008} implies that the $u^{-2}$-diffusion equation
admits an infinite series $\{\mathfrak g_p,\, p\in\mathbb N\}$
of potential symmetry algebras.
For any $p\in\mathbb N$ the algebra $\mathfrak g_p$ is of strictly $p$th potential order and is
associated with $p$-tuples of the linearly independent lowest degree polynomial
solutions of the backward heat equation.
Moreover, each algebra $\mathfrak g_p$ is isomorphic to the maximal Lie invariance algebra of the linear heat equation.
The case 3.1 is a specification of the above general frame, corresponding to the pair $(1,x)$ of the simplest solutions
of the linear heat equation.

If $f=x^{-4/3}$ and $n=-2$, system~\eqref{EqTransformedPotSysForEqDifEqf} is equivalent
to the $\tilde v^{-4/3}$-diffusion equation $\tilde v_{\tilde t}=(\tilde v^{-4/3}\tilde v_{\tilde x})_{\tilde x}$.
The power nonlinearity of degree $-4/3$ is a well-known case of Lie symmetry extension from the Ovsiannikov's group classification of
nonlinear diffusion equations~\cite{Ovsiannikov1982}.
Each Lie symmetry of system~\eqref{EqTransformedPotSysForEqDifEqf} is the first-order prolongation, with respect to~$x$,
of a Lie symmetry of the $\tilde v^{-4/3}$-diffusion equation.
Each Lie symmetry of the $\tilde v^{-4/3}$-diffusion equation is prolonged to a Lie symmetry of system~\eqref{EqTransformedPotSysForEqDifEqf}
and then mapped, by transformation~\eqref{EqHodographXV}, to a Lie symmetry of the system~{\bf1} with $f=x^{-4/3}$ and $n=-2$.
Therefore, the Lie invariance algebra of case 1.2 is isomorphic to the Lie invariance algebra of the $\tilde v^{-4/3}$-diffusion equation.

If $f=1$ and $n=-2/3$, system~\eqref{EqTransformedPotSysForEqDifEqf} coincides with the potential system
of the $\tilde u^{-4/3}$-diffusion equation $\tilde u_{\tilde t}=(\tilde u^{-4/3}\tilde u_{\tilde x})_{\tilde x}$,
associated with the characteristic~1.
The nontrivial Lie symmetry operator $x^2\p_x-3xu\p_u$ of this equation cannot be prolonged to a single potential.
This is why the corresponding case of potential symmetry of equations from class~\eqref{eqDifEqf}
arises only under consideration of the united potential system~{\bf3} having two potential variables.

\section{A class of porous medium equations}\label{SectionPorMedEq}

The second class examined in~\cite{Gandarias2008} consists of porous medium equations having the form
\begin{equation}\label{eqPorMedEq2}
u_t=(u^n)_{xx}+h(x)(u^m)_x.
\end{equation}
Here $f(x)/m$ from~\cite{Gandarias2008} is redenoted by $h(x)$ for convenience.
Only the case $m=n$ was considered in~\cite{Gandarias2008} but this investigation in fact is needless.
The general principle of group analysis of differential equations is that objects
(differential equations, classes of differential equations, exact solutions,
subalgebras of Lie invariance algebras etc.) are assumed \emph{similar} if they are related via point transformations
\cite{Olver1993,Ovsiannikov1982,Popovych&Kunzinger&Eshraghi2006,VPS2008}.
Such objects have similar transformational or other properties relevant to the framework of group analysis
and hence a single representative from a set of similar objects is enough to be investigated.
Any consideration in another style must be additionally justified.
The subclass of class~\eqref{eqPorMedEq2}, singled out by the condition $m=n$, is mapped to class~\eqref{eqDifEqf}
via a family of point transformations parameterized by the arbitrary element~$h$.
Indeed, any equation of form~\eqref{eqPorMedEq2} is reduced
by the transformation
\begin{equation}\label{EqTransOfEqPorMedEq2ToEqDifEqf}
\tilde t=t, \quad \tilde x=\int\! e^{-\int\!h\,dx}dx,\quad \tilde u=u,
\end{equation}
to the equation of form~\eqref{eqDifEqf} with $f(\tilde x)=e^{2\int\!h\,dx}/n$ and $\tilde n=n-1$.
Transformation~\eqref{EqTransOfEqPorMedEq2ToEqDifEqf} was found in~\cite{Ivanova&Popovych&Sophocleous2004,Ivanova&Popovych&Sophocleous2006Part1}
as an element of the generalized extended equivalence group of the wider class of
nonlinear variable-coefficient diffusion--convection equations of the general form
\[f(x)u_t=(A(u)u_x)_x + h(x)B(u)u_x,\] where $fA\ne0$.
(The additional attribute ``extended'' means that transformations from the group may depend on
arbitrary elements of the class in a nonlocal way.)
The properties of possessing conservation laws and potential symmetries
completely agree with the similarity relation of differential equations
\cite{Popovych&Ivanova2004ConsLawsLanl,Popovych&Kunzinger&Ivanova2008}.
Hence any result on potential symmetries of equations from class~\eqref{eqPorMedEq2}
can be easily derived via the application of the inverse to transformation~\eqref{EqTransOfEqPorMedEq2ToEqDifEqf}
as a reformulation of the corresponding result for class~\eqref{eqDifEqf}.
See the discussion on potential symmetries of equations from class~\eqref{eqDifEqf} in Section~\ref{SectionDifEq}.
Thus, potential symmetries of~\eqref{eqPorMedEq2} is found in~\cite{Gandarias2008} only in the case $h=-1/(2x)$
(formula~(36) of~\cite{Gandarias2008}).
All of them are the preimages of symmetries presented in case 2.2 of Section~\ref{SectionDifEq}, where $c_1$ has to be set to equal 0.
In other words, more general results on potential symmetries of equations from class~\eqref{eqPorMedEq2}
than presented in~\cite{Gandarias2008} are in fact known.

\section{A second class of porous medium equations}\label{SectionPorMedEq2}

In~\cite{Gandarias2008} the porous medium equations of the form
\begin{equation}\label{eqPorMedEq}
u_t=\left((u^n)_x+f(x)u^m\right)_x,
\end{equation}
with $n\ne0$ was given without considering its potential symmetries.
It was stated that the complete classification of potential symmetries was carried out in~\cite{Gandarias1996}.
There are three remarks on this statement.

1) In~\cite{Gandarias1996} the case $m=0$ was omitted from the consideration since
another representation for which the value $m=0$ is singular was used for equations from class~\eqref{eqPorMedEq}.
At the same time, the corresponding subclass of class~\eqref{eqPorMedEq} contains well-known equations,
e.g., the linearizable equation $u_t=(u^{-2}u_x)_x+1$.
Moreover, some equations from the cases $m=0$ and $m\ne0$ are connected within both the point and potential frame.

2) The description of potential symmetries in~\cite{Gandarias1996} was not a classification
since no equivalence relations of equations or symmetries were used.

3) Only simplest potential symmetries arising under the study of the corresponding ``natural'' potential systems were found.
The problem on the construction of the other simplest and, moreover, general potential symmetries of equations from class~\eqref{eqPorMedEq}
was still open.

We employ class~\eqref{eqPorMedEq} in order to give the
basic steps for the exhaustive classification of the simplest potential symmetries.
Moreover, in the next section we completely describe potential symmetries of some equations from class~\eqref{eqPorMedEq}.

The equivalence group $G^\sim$ of class~\eqref{eqPorMedEq} is formed by the transformations
\begin{gather*}\tilde t=\delta_1t+\delta_2,\quad\tilde x=\delta_3x+\delta_4,\quad\tilde
u=\delta_1{}^{-\frac1{n-1}}\delta_3{}^{\frac2{n-1}}u,\\
\tilde f=
\delta_1{}^{\frac{m-n}{n-1}}\delta_3{}^{\frac{n-2m+1}{n-1}}f,\quad\tilde n=n,\quad\tilde m=m,\end{gather*}
where $\delta_i,$  $i=1,\dots,4,$ are arbitrary constants,
$\delta_1\delta_3\ne0.$
Additionally, the subclass singled out from~\eqref{eqPorMedEq} by the condition $m=n$
can be mapped to the subclass consisting of the equations
\begin{equation*}\label{eqDif_f_exp}
\quad e^{-\frac {n+1}n\int f(x) dx}\tilde u_{\tilde t}=(\tilde u{}^{n})_{\tilde x\tilde x}
\end{equation*}
by the transformation
\begin{equation}\label{transPorMedToInhDifEq}
\tilde t=t,\quad\tilde x=\int e^{\int f(x) dx}dx,\quad\tilde u=e^{\frac1n\int f(x) dx}u.
\end{equation}

We present the conservation laws for (\ref{eqPorMedEq}) and the subsequent potential systems.

\begin{theorem}\label{TheoremCLofPorMedEq}
Any equation from class~\eqref{eqPorMedEq} has the conservation law of form~\eqref{conslaw}
whose density and flux are, respectively,
\begin{equation}\textstyle\label{ker.cons.law}
1.\quad F=u,\qquad G=-nu^{n-1}u_x-fu^m.
\end{equation}
A complete list of $G^{\Equiv}$-inequivalent equations~\eqref{eqPorMedEq} having
additional (i.e. linear independent with~\eqref{ker.cons.law}) conservation laws
is exhausted by the following ones
\begin{gather*}\textstyle
2.\quad m=n\neq1: \qquad  F=u\int\!e^{\int\! fdx}dx, \quad G=-\int\!e^{\int\! fdx}dx\,(nu^{n-1}u_x+fu^n)+e^{\int\! fdx}u^n,
\\[.5ex]\textstyle
3.\quad n\neq1, \quad m=0: \qquad F=xu, \quad G=-x(nu^{n-1}u_x+f)+u^n+\int fdx,
\\[.5ex] \textstyle
4.\quad n\neq1, \quad m=1,\quad f=1:
 \qquad F=(t+x)u, \quad G=-(t+x)(nu^{n-1}u_x+u)+u^n,  \\[.5ex]\textstyle
5.\quad n\neq1, \quad m=1,\quad f={\varepsilon}x : \qquad F=e^{\varepsilon t}xu,
\quad G=-e^{\varepsilon t}x(nu^{n-1}u_x+xu)+e^{\varepsilon t}u^n,
\\[.5ex] \textstyle
6.\quad n=1,\quad m=0: \qquad F=\alpha u, \quad G=-\alpha(u_x+f)+\alpha_xu+\int\!\alpha_xfdx,
\\[.5ex] \textstyle
7.\quad n=1,\quad m=1: \qquad F=\beta u, \quad G=-\beta(u_x+fu)+\beta_xu,
\end{gather*}
where $\varepsilon=\pm1\bmod G^{\Equiv},$
$\alpha=\alpha(t,x)$ and $\beta=\beta(t,x)$ are arbitrary solutions of the linear equations $\alpha_t+\alpha_{xx}=0$
and $\beta_t+\beta_{xx}-f\beta_x=0$, respectively.
(Together with restrictions on values $f$, $n$ and $m$ we also adduce densities and fluxes of additional conservation laws.)
\end{theorem}

These conservation laws can be used for the construction of potential systems that lead to potential symmetries
for the equation (\ref{eqPorMedEq}).
The associated characteristics are equal to the coefficients of~$u$ in the presented expressions for~$F$.

Here we consider only simplest potential systems
(i.e., potential systems with one potential variable, constructed
with usage of single conserved vectors of basis conservation laws) of
equations from class~\eqref{eqPorMedEq}, having the form
\[
v_x=F, \quad v_t=-G.
\]
Cases~6 and~7 of Theorem~\ref{TheoremCLofPorMedEq} can be excluded from
the investigation since they concern linear equations studied in~\cite{Popovych&Kunzinger&Ivanova2008}.
Then, the equations of case~2 are reducible to diffusion equations of form~\eqref{eqDifEqf}
by transformation~\eqref{transPorMedToInhDifEq}.
The equations of cases 4 and 5 are reducible to
the constant coefficient diffusion equation~$\tilde u_{\tilde t}=(\tilde u{}^{n})_{\tilde x\tilde x}$
by means of the Galilei transformation
\[
\tilde t=t, \quad \tilde x=x+t,\quad \tilde u=u
\]
and the transformation
\[\tilde t=\begin{cases}\frac1{\varepsilon (n+1)}e^{\varepsilon(n+1) t},& n\not=-1,\\t ,& n=-1,
\end{cases}
\quad\tilde x=e^{\varepsilon t}x,\quad\tilde u=e^{-\varepsilon t}u,\]
respectively.

Therefore, we have to investigate only two potential systems
\begin{gather}
v_x=u, \quad v_t=nu^{n-1}u_x+fu^m ,\label{sysPotSys1PorMedEq}\\ \textstyle
v^*_x=xu, \quad v^*_t=x(nu^{n-1}u_x+f)-u^n-\int fdx \label{sysPotSys3PorMedEq}
\end{gather}
corresponding to cases~1 and~3 of Theorem~\ref{TheoremCLofPorMedEq}.
(To distinguish the potential introduced, we denote the second potential by~$v^*$.)

Lie point symmetries of the potential system~\eqref{sysPotSys1PorMedEq}
give nontrivial potential symmetries of equation~\eqref{eqPorMedEq} in the following cases:

\medskip

\noindent1.\quad $f=x^{-2}$, $n=-1$, $m=-2$: \\[.5ex]
\phantom{1.\quad}$\langle\p_t,\, \p_{v},\, x\p_x-u\p_u,\, 12t\p_t+(3\ln x-v)x\p_x+
(3-3\ln x+xu+v)u\p_u+2(3v-t)\p_v\rangle$;

\medskip

\noindent2.\quad $f=\varepsilon x\ln |x|$, $\varepsilon=\pm1$, $n={-1}$, $m=1$:\\[.5ex]
\phantom{2.\quad}$\langle\p_t,\; \p_v,\; e^{-\varepsilon t}(x\p_x-u\p_u),\;
e^{-\varepsilon t}(-\varepsilon xv\p_x+\varepsilon(xu^2+uv)\p_u+2\p_v)\rangle$;
\medskip

\noindent3.\quad $f=x$, $n=-1$, $m=0$: \\[.5ex]
\phantom{3.\quad}$\langle\p_t,\, \p_{v},\, \p_x+t\p_v,\,
2t\p_t-x\p_x+2u\p_u+v\p_v,\,
t^2\p_t+(v-tx)\p_x+(2t-u)u\p_u+tv\p_v\rangle$;

\medskip

\noindent4.\quad $f=0$, $n={-1}$, $m=0$:\\[.5ex]
\phantom{4.\quad}$\langle\p_t,\; \p_v,\; x\p_x-u\p_u,\;2t\p_t+u\p_u+v\p_v,\; vx\p_x-(v+xu)u\p_u+2t\p_v,$\\[.5ex]
\phantom{4.\quad}$4t^2\p_t+(v^2-2t)x\p_x-(v^2+2vxu-6t)u\p_u+4tv\p_v,\;\beta\p_x-\beta_vu^2\p_u\rangle$;

\medskip

\noindent5.\quad $f=1$, $n={-1}$, $m=1$:\\[.5ex]
\phantom{5.\quad}$\langle\p_t,\; \p_v,\; (x+t)\p_x-u\p_u,\; 2t\p_t+2x\p_x-u\p_u+v\p_v,\; v(x+t)\p_x-[(x+t)u+v]u\p_u+2t\p_v,$\\[.5ex]
\phantom{5.\quad}$4t^2\p_t+[(x+t)v^2-2tx-6t^2]\p_x-[v^2-6t+2(x+t)vu]u\p_u+4tv\p_v,\;\beta\p_x-\beta_vu^2\p_u\rangle$;

\medskip

\noindent6.\quad $f=\varepsilon x$, $\varepsilon=\pm1$, $n={-1}$, $m=1$:\\[.5ex]
\phantom{6.\quad}$\langle\p_t,\; \p_v,\; x\p_x-u\p_u,\; 2t\p_t-2\varepsilon tx\p_x+(1+2\varepsilon t)u\p_u+v\p_v,\;
vx\p_x-(xu+v)u\p_u+2t\p_v,$\\[.5ex]
\phantom{6.\quad}$4t^2\p_t+\left(v^2-2t-4\varepsilon t^2\right)x\p_x-\left(v^2-6t+2xvu-4\varepsilon t^2\right)u\p_u+4tv\p_v,\;
e^{-\varepsilon t}(\beta\p_x-\beta_vu^2\p_u)\rangle$;

\medskip

\noindent7.\quad $n={-1}$, $m=-1$:\\[.5ex]
\phantom{7.\quad}$\langle\p_t,\; \p_v,\; \psi\p_x-(1-f\psi)u\p_u,\;  2t\p_t+u\p_u+v\p_v,\; \psi v\p_x-((1-f\psi)v+\psi u)u\p_u+2t\p_v,$ \\[.5ex]
\phantom{7.\quad}$4t^2\p_t+(v^2-2t)\psi\p_x-[2\psi vu+(v^2-2t)(1-f\psi)-4t]u\p_u+4tv\p_v,$ $\varphi(\beta \p_x-(\beta_vu^2-f\beta u)\p_u)\rangle;$.

\medskip

\noindent8.\quad $f=1$, $n=1$, $m=2$:\\[.5ex]
\phantom{8.\quad}$\langle\p_t,\; \p_x,\; \p_v,\; 2t\p_t+x\p_x-u\p_u,\; 2t\p_x-\p_u-x\p_v,\; 4t^2\p_t+4tx\p_x-2(2tu+x)\p_u-(2t+x^2)\p_v,$\\[.5ex]
\phantom{8.\quad}$e^{-v}[(\alpha u-\alpha_x)\p_u-\alpha\p_v]\rangle$.

\medskip

Here $\alpha=\alpha(t,x)$ and $\beta=\beta(t,v)$ run through the solution sets of the linear heat equation
$\alpha_t-\alpha_{xx}=0$ and backward linear heat equation $\beta_t+\beta_{vv}=0$,
respectively, $\varphi(x)=e^{-\int\! fdx},$ $\psi(x)=e^{-\int\! fdx}\int\! e^{\int\! fdx}dx.$

\begin{note}
Equations from class~\eqref{eqPorMedEq} with $n={-1}$, $m=0$ and  $f\in\{0,1\}$  are just different representations of the same equation.
Potential systems corresponding to these two cases are connected via the transformation $\tilde v=v+t$ of potential variable $v$.
This transformation maps case~4 to the case

\medskip

\noindent\phantom{8.\quad}
$f=1$, $n={-1}$, $m=0$:
\\[.5ex]\phantom{8.\quad}
$\langle\p_t,\; \p_{\tilde v},\; x\p_x-u\p_u,\;  2t\p_t+u\p_u+(t+{\tilde v})\p_{\tilde v},\;
({\tilde v}-t)x\p_x-({\tilde v}-t+xu)u\p_u+2t\p_{\tilde v},$
\\[.5ex]\phantom{8.\quad}
$4t^2\p_t+[({\tilde v}-t)^2-2t]x\p_x-[({\tilde v}-t)^2+2({\tilde v}-t)xu-6t]u\p_u+4t{\tilde v}\p_{\tilde v},$
\\[.5ex]\phantom{8.\quad}
$e^{\frac t4-\frac {\tilde v}2}[\tilde\beta\p_x-(\tilde\beta_{\tilde v}-\frac12\tilde\beta)u^2\p_u]\rangle$,

\medskip

\noindent
where the function $\tilde\beta=\tilde\beta(t,\tilde v)$ runs through the solution set of the backward linear heat equation
$\tilde\beta_t+\tilde\beta_{\tilde v\tilde v}=0$.
\end{note}

\begin{note}
Cases 5, 6 and 7 are reduced to  case 4 by the point transformations $\{\tilde x=x+t,\,\tilde u=u\}$,
$\{\tilde x=e^{\varepsilon t}x,\,\tilde u=e^{-\varepsilon t}u\}$ and~\eqref{transPorMedToInhDifEq}, respectively.
The variables~$t$ and~$v$ are identically transformed.
\end{note}

As mentioned, Lie symmetries of potential system~\eqref{sysPotSys1PorMedEq} were investigated in~\cite{Gandarias1996}
only for $m\ne0$ and hence cases~3 and~4 were omitted there.
It is explained by choice of another representation of equation~\eqref{eqPorMedEq}.
For all values of~$m$, potential symmetries of equation~\eqref{eqPorMedEq}
associated with potential system~\eqref{sysPotSys1PorMedEq} are first classified above.

There exists only one inequivalent case of potential system~\eqref{sysPotSys3PorMedEq}
that gives nontrivial potential symmetries for equation~\eqref{eqPorMedEq},
namely, $f=x$, $n=-1$ and $m=0$. Lie algebra of potential symmetries in this case has the form

\medskip

\noindent9.\quad $f=x$, $n=-1$, $m=0$:\\[.5ex]
\phantom{9.\quad}$\langle \p_t,\; \p_v,\; 2t\p_t-x\p_x+2u\p_u,\; e^{-v^*/2}(2x\p_x+(x^2u-2)u\p_u-4\p_{v^*}) \rangle.$

\begin{note}
In this section we have first presented the classification of local conservation
laws and the simplest potential symmetries of equations from class~\eqref{eqPorMedEq}.
To complete investigation of potential symmetries in this class,
it is necessary to study potential systems depending on several potentials and systems
constructed with potential conservation laws of~\eqref{eqPorMedEq}.
Using the potential equivalence methods, in the next section we describe general potential symmetries
of linearized equations from this class.
\end{note}

Here we give only one example on potential symmetries of equations from class~\eqref{eqPorMedEq} with $f\ne0$,
which involve two potentials.
They are easily constructed via the point transformation~\eqref{transPorMedToInhDifEq}
from potential symmetries presented in Section~\ref{SectionDifEq}.
Namely, the equation $x^{-6}u_t=(u^{-2/3}u_x)_x$ from class~\eqref{eqDifEqf} (case 3.4, $f=x^{-6}$, $n={-2/3}$)
is mapped by the transformation
$\tilde t =\frac34t$,
$\tilde x =|x|^{-1/2}$,
$\tilde u =|x|^{-9/2}u$
to the equation
\begin{equation}\label{EqPorMedEq1313-3x}
\tilde u_{\tilde t}=((\tilde u^{1/3})_{\tilde x}-3\tilde x^{-1}\tilde u^{1/3})_{\tilde x}
\end{equation}
from class~\eqref{eqPorMedEq}, where $\tilde n=\tilde m=1/3$ and $\tilde f=-3\tilde x^{-1}$.
For coefficients to be simpler, we have additionally combined the corresponding transformation of form~\eqref{transPorMedToInhDifEq}
with a scaling.
The second order potential system for equation~\eqref{EqPorMedEq1313-3x},
which is constructed from~\eqref{SystemPotSysSecondLevel} via the transformation prolonged to the potentials as
$\tilde v=-v^1/2$ and $\tilde w=w/4$, is
\[
\tilde v_{\tilde x}=\tilde u, \quad
\tilde w_{\tilde x}=\tilde x^{-3}\tilde v, \quad
\tilde w_{\tilde t}=\tilde x^{-3}\tilde u^{1/3}.
\]
The associated potential symmetry algebra of~\eqref{EqPorMedEq1313-3x}  is (we omit tildes for convenience)
\begin{gather*}
\langle\p_t,\ \p_w,\ 2\p_v-x^2\p_w,\ 2t\p_t+3u\p_u+3v\p_v+3w\p_w,\ x\p_x-3u\p_u-2v\p_v-4w\p_w,
\\
x^{-1}\p_x+3x^{-2}u\p_u+2(2w+x^{-2}v)\p_v-2x^{-2}w\p_w,
\\
xw\p_x-3(x^{-2}v+w)u\p_u-(x^{-2}v+2w)v\p_v-2w^2\p_w\rangle.
\end{gather*}

\section{Applications of potential equivalence transformations}%
\label{SectionOnApplicationsOfPETs}

In a way analogous to class~\eqref{eqDifEqf} (see the end of Section~\ref{SectionDifEq}),
potential equivalence transformations (PETs) can be effectively applied to explain the provenance of
potential symmetries of equations from class~\eqref{eqPorMedEq}.
The main tool again is the potential hodograph transformation
\begin{equation}\label{EqHodographXV2}
\tilde t=-t,\quad \tilde x=v,\quad\tilde u=u^{-1},\quad \tilde v=x.
\end{equation}
Transformation~\eqref{EqHodographXV2} differs from~\eqref{EqHodographXV} in the sign of~$t$
due to a difference in the representations of classes~\eqref{eqDifEqf} and~\eqref{eqPorMedEq}
and will be additionally modified via composing with point transformations
in order to present the imaged equations in canonical forms.
We need to interpret only cases 1--3, 8 and 9 of nontrivial potential symmetries from Section~\ref{SectionPorMedEq2}
since cases~5--7 are reduced to case 4 by point transformations
and case 4 coincides, up to alternating the sign of~$t$, with case 1.1 from Section~\ref{SectionDifEq}
whose interpretation is presented in the end of that section.
We also consider subclasses including cases to be interpreted and show
that the provenance of potential symmetries is connected with extensions of Lie symmetry groups of
the corresponding potential equations.\looseness=-1

\medskip

{\bf\mathversion{bold}$f=1$, $n=1$, $m=2$} (case 8).
The equation of form~\eqref{eqPorMedEq} with these values of the arbitrary elements is the Burgers equation $u_t=u_{xx}+2uu_x$.
The associated potential system $v_x=u$, $v_t=u_x+u^2$
is mapped to the potential system $\tilde v_{\tilde x}=\tilde u$, $\tilde v_{\tilde t}=\tilde u_{\tilde x}$
of the linear heat equation~$\tilde u_{\tilde t}=\tilde u_{\tilde x\tilde x}$,
constructed with the conservation law having the characteristic 1,
by the transformation~$\mathcal T$: $\tilde t=t$,  $\tilde x=x$, $\tilde u=u e^v$, $\tilde v=e^v$.
Analogously to the $u^{-2}$-diffusion equation,
the whole set of potential systems (of all levels) of the Burgers equation consists of
the preimages of the potential systems of the linear heat equation, corresponding to the tuples of characteristics
including the characteristic~1 as the first element, with respect to transformation~$\mathcal T$
identically prolonged to the other potentials.
Let us recall that the potential systems of the linear heat equation
are exhausted by the first-level potential systems of form~{\bf4} with~$f=1$ (see section~\ref{SectionDifEq})
\cite{Popovych&Ivanova2004ConsLawsLanl,Popovych&Kunzinger&Ivanova2008}.
Therefore, the maximal level of potentials and potential systems of the Burgers equation equals two.
All properties of potential symmetries of the Burgers equation follow from the similar properties
of the linear heat equation, including the existence of an infinite series of potential symmetry algebras
of all possible potential orders, which are isomorphic to the maximal Lie invariance algebra of the linear heat equation.

\medskip

{\bf\mathversion{bold}$n=-1$, $m=0$.}
We have the mapping
\[\hspace*{-.5\arraycolsep}\begin{array}{lll}
v_x=u, \  v_t=-u^{-2}u_x+f&\sim&u_t=(u^{-1})_{xx}+f_x\\
\bigg\downarrow\ \eqref{EqHodographXV2}&&\\
\tilde v_{\tilde x}=\tilde u,\ \tilde v_{\tilde t}={\tilde u}_{\tilde x}+f\tilde u&\sim&
\tilde v_{\tilde t}=\tilde v_{\tilde x\tilde x}+f(\tilde v)\tilde v_{\tilde x}.
\end{array}\]
The imaged `potential' equations form the class of semilinear convection--diffusion equations
whose group classification is known~\cite{Popovych&Ivanova2005PETs}.
Only two inequivalent cases of this classification (the linear heat equation with $f=0$ and the Burgers equation with $f=\tilde v$)
lead to inequivalent cases of potential symmetries in class~\eqref{eqPorMedEq}.
We neglect the case $f=0$ as was already discussed in Sections~\ref{SectionFokPlEq} and~\ref{SectionDifEq}.

In the case $f=x=\tilde v$ both  potential systems~\eqref{sysPotSys1PorMedEq} and~\eqref{sysPotSys3PorMedEq}
give nontrivial potential symmetries.
The corresponding equation from class~\eqref{eqPorMedEq} is a well known integrable equation~\cite[p.~129]{Mikhailov&Shabat&Sokolov1991}
(see also \cite[p.~328]{Olver1993}).

The maximal Lie symmetry algebra of the potential system~\eqref{sysPotSys1PorMedEq} (case~3)
is the preimage, with respect to the transformation
\[
\hspace*{-.5\arraycolsep}\begin{array}{lll}
v_x=u, \  v_t=-u^{-2}u_x+x&\sim&u_t=(u^{-1})_{xx}+1\\
\bigg\downarrow\ \eqref{EqHodographXV2}&&\\
\tilde v_{\tilde x}=\tilde u,\ \tilde v_{\tilde t}=\tilde u{}_{\tilde x}+\tilde v\tilde v_{\tilde x}&\sim&
\tilde v_{\tilde t}=\tilde v_{\tilde x\tilde x}+\tilde v\tilde v_{\tilde x},
\end{array}
\]
of the maximal Lie symmetry algebra
\[
\langle\p_{\tilde t},\; \p_{\tilde x},\; \tilde t\p_{\tilde x}-\p_{\tilde v},\;
2\tilde t\p_{\tilde t}+\tilde x\p_{\tilde x}-\tilde v\p_{\tilde v},\;
{\tilde t}^2\p_{\tilde t}+\tilde t\tilde x\p_{\tilde x}-(\tilde t\tilde v+\tilde x)\p_{\tilde v} \rangle
\]
of the Burgers equation $\tilde v_{\tilde t}=\tilde v_{\tilde x\tilde x}+\tilde v\tilde v_{\tilde x}$,
prolonged to~$\tilde u$ according to the equality $\tilde u=\tilde v_{\tilde x}$.
Therefore, these algebras are isomorphic.

For the potential system~\eqref{sysPotSys3PorMedEq} with $f=x$, $n=-1$ and $m=0$ (case 9)
we need to modify transformation~\eqref{EqHodographXV2}:
\[
\hspace*{-.5\arraycolsep}\begin{array}{lll}\displaystyle
v^*_x=xu,\  v^*_t=-xu^{-2}u_x-u^{-1}+\frac12x^2&\sim&u_t=(u^{-1})_{xx}+x\\[1ex] \displaystyle
\Bigg\downarrow\ \tilde t=-\frac t4,\  \tilde x=\frac {v^*}2,\ \tilde u=\frac2u,\  \tilde v=\frac1x&&\\[2ex]
\tilde v_{\tilde x}=-\tilde v^3\tilde u,\ \tilde v_{\tilde t}=-(\tilde v\tilde u)_{\tilde x}
+\tilde v^{-2}\tilde v_{\tilde x}&\sim&\tilde v_{\tilde t}=(\tilde v^{-2}\tilde v_{\tilde x})_{\tilde x}+\tilde v^{-2}\tilde v_{\tilde x}
\end{array}
\]
The equation $\tilde v_{\tilde t}=(\tilde v{}^{-2}\tilde v_{\tilde x})_{\tilde x}+\tilde v{}^{-2}\tilde v_{\tilde x}$
arises under the group classification of convection--diffusion equations~\cite{Popovych&Ivanova2005PETs}.
It is reduced to the remarkable diffusion equation $\hat v_{\hat t}=(\hat v{}^{-2}\hat v_{\hat x})_{\hat x}$
by the point transformation $\hat t=\tilde t$, $\hat x=e^{\tilde x}$, $\hat v=e^{-\tilde x}\tilde v$.
Its maximal Lie invariance algebra is
\[
\mathcal A=
\langle\p_{\tilde t},\;\p_{\tilde x},\;2\tilde t\p_{\tilde t}+\tilde v\p_{\tilde v},\;
e^{-\tilde x}\p_{\tilde x}+e^{-\tilde x}\tilde v\p_{\tilde v}\rangle.
\]
The prolongation of~$\mathcal A$ to~$\tilde u$ according to the equality $\tilde u=-\tilde v^{-3}\tilde v_{\tilde x}$ is an image of
the algebra of case~9.
Therefore, the algebra of case~9 is isomorphic to $\mathcal A$ and the maximal Lie invariance algebra of the diffusion equation
$\hat v_{\hat t}=(\hat v{}^{-2}\hat v_{\hat x})_{\hat x}$.

The united potential system of~\eqref{sysPotSys1PorMedEq} and~\eqref{sysPotSys3PorMedEq} with $f=x$, $n=-1$ and $m=0$ also admits Lie symmetries
inducing potential symmetries of the equation $u_t=(u^{-1})_{xx}+1$.
It is mapped to a system equivalent to the linear heat equation by transformation~\eqref{EqHodographXV2}
supplemented with $\tilde v^*=e^{v^*/2}$.
Namely, we have the transformation
\[
\hspace*{-.5\arraycolsep}\begin{array}{lll}
v_x=u, \  v_t=-u^{-2}u_x+x,\ v^*_x=xu,\  v^*_t=-xu^{-2}u_x-u^{-1}+\frac12x^2&\sim&u_t=(u^{-1})_{xx}+1\\
\bigg\downarrow\ \eqref{EqHodographXV2},\ \tilde v^*=e^{v^*/2}&&\\
\tilde v_{\tilde x}=\tilde u,\ \tilde v_{\tilde t}=\tilde u{}_{\tilde x}+\tilde v\tilde v_{\tilde x},\
\tilde v^*_{\tilde x}=\frac12\tilde v^*\tilde v,\ \tilde v^*_{\tilde t}=\frac14\tilde v^*\tilde v^2+\frac12\tilde v^*\tilde u
&\sim&
\tilde v^*_{\tilde t}=\tilde v^*_{\tilde x\tilde x}.
\end{array}
\]
As a result,
the whole set of inequivalent potential systems (of all levels) of the equation  $u_t=(u^{-1})_{xx}+1$ consists of
systems~\eqref{sysPotSys1PorMedEq} and~\eqref{sysPotSys3PorMedEq}, the united system of~\eqref{sysPotSys1PorMedEq} and~\eqref{sysPotSys3PorMedEq}
and the systems obtained by the following procedure:
We take any potential system of the linear heat equation $\tilde v^*_{\tilde t}=\tilde v^*_{\tilde x\tilde x}$
and supplement it with the equations $\tilde v=2\tilde v^*_{\tilde x}/\tilde v^*$ and $\tilde u=\tilde v_{\tilde x}$
defining $\tilde v$ and $\tilde u$.
The equations $\tilde v_{\tilde t}=\tilde u_{\tilde x}+\tilde v\tilde v_{\tilde x}$ and
$\tilde v^*_{\tilde t}=\frac14\tilde v^*\tilde v^2+\frac12\tilde v^*\tilde u$ are differential consequences of
the above equations in $\tilde v^*$, $\tilde v$ and $\tilde u$.
Then the inverse of transformation~\eqref{EqHodographXV2} extended to~$v^*$ by $\tilde v^*=e^{v^*/2}$ and identically
extended to the other potentials gives the potential system of the equation $u_t=(u^{-1})_{xx}+1$.

The studied structure of the set of potential symmetries allows us to conclude that
the potential symmetries of this equation are exhausted by cases 3 and~9 and the potential symmetry algebras
constructed from Lie and potential symmetry algebras of the linear heat equation using the described procedure
of extension and mapping.

Finally, the above consideration shows that the well known linear and linearizable equations
$u_t=u_{xx}$, $u_t=(u^{-2}u_x)_x$, $u_t=(u^{-2}u_x)_x+u^{-2}u_x$, $u_t=u_{xx}+uu_x$ and $u_t=(u^{-2}u_x)_x+1$
are singled out from classes of quasilinear second-order evolution equations with the properties of possessing
infinite series of infinite-dimensional algebras of potential symmetries, isomorphic to
the maximal Lie invariance algebra of the linear heat equation.
Moreover, these equations are connected to each other by potential equivalence transformations.

\medskip

{\bf\mathversion{bold}$n=-1$, $m=1$.}
The corresponding subclass of class~\eqref{eqPorMedEq} is mapped by~\eqref{EqHodographXV2} in the following way
\[\hspace*{-.5\arraycolsep}\begin{array}{lll}
v_x=u, \  v_t=-u^{-2}u_x+fu&\sim&u_t=((u^{-1})_x+fu)_x\\
\bigg\downarrow\ \eqref{EqHodographXV2}&&\\
\tilde v_{\tilde x}=\tilde u,\ \tilde v_{\tilde t}=\tilde u_{\tilde x}+f&\sim&\tilde v_{\tilde t}={\tilde v}_{\tilde x\tilde x}+f(\tilde v)
\end{array}\]
The imaged `potential' equations
\begin{equation}\label{EqNHEinV}
\tilde v_{\tilde t}={\tilde v}_{\tilde x\tilde x}+f(\tilde v)
\end{equation}
form the class of semilinear heat equations with sources.
The problem of group classification in class~\eqref{EqNHEinV} was solved in~\cite{Dorodnitsyn1979}
(see also~\cite{Ibragimov1994V1}).
The linear equations from the list presented in~\cite{Dorodnitsyn1979} induce cases 4--6 already discussed.
Excluding them, there is only one representative in the list, whose preimage with respect to~\eqref{EqHodographXV2} possesses potential symmetries.
Namely, the equation of form~\eqref{EqNHEinV} with $f=\varepsilon\,\tilde v\ln\tilde v$ has
the maximal Lie invariance algebra
\[
\langle\p_{\tilde t},\ \p_{\tilde x},\ e^{\varepsilon\tilde t}\tilde v\p_{\tilde v},\
2e^{\varepsilon\tilde t}\p_{\tilde x}-\varepsilon e^{\varepsilon\tilde t}\tilde x\tilde v\p_{\tilde v}\rangle.
\]
Preimaging this case of Lie symmetry extension in the class~\eqref{EqNHEinV},
we obtain case 2 of the  classification of potential symmetries in class~\eqref{eqPorMedEq}.

The same trick with preimaging with respect to the potential hodograph transformation~\eqref{EqHodographXV2}
can be applied to construct potential nonclassical symmetries in the initial subclass singled out by the conditions $n=-1$ and $m=1$.
This is easy to do because nonclassical symmetries of the equations from the class~\eqref{EqNHEinV} have been already investigated
\cite{ArrigoHillBroadbridge1993,Clarkson&Mansfield1993,FushchichSerov1990,VPS2008}.
The corresponding exact solutions were constructed by the reduction method
in~\cite{ArrigoHillBroadbridge1993,Clarkson&Mansfield1993}, see also~\cite{VPS2008}.
Nonlinear equations of form~\eqref{EqNHEinV} possess
purely nonclassical symmetries with nonvanishing coefficients of $\partial_{\tilde t}$
if and only if $f$ is a cubic polynomial in $\tilde v$, i.e., $f=a\tilde v^3+b\tilde v^2+c\tilde v+d$,
where $a$, $b$, $c$, and $d$ are arbitrary constants, $a\neq0$
and the coefficient~$b$ can be put equal to~0 by translations with respect to~$v$.
Up to the equivalence generated by translations with respect to~$v$ and~$x$,
such symmetries are exhausted by the following operators (hereafter  $b=0$ and $\mu=\sqrt{|c|/2}$):
\[\arraycolsep=0ex
\begin{array}{ll}
a<0\colon\quad&
\partial_{\tilde t}\pm\frac32\sqrt{-2a}\,{\tilde v}\partial_{\tilde x}+\frac32(a {\tilde v}^3+c {\tilde v}+d)\partial_{\tilde v},\\[1ex]
c=0,\,d=0\colon\quad&
\partial_{\tilde t}-3x^{-1}\partial_{\tilde x}-3x^{-2}{\tilde v}\partial_{\tilde v},\\[1ex]
c<0,\,d=0\colon\quad&
\partial_{\tilde t}+3\mu\tan(\mu {\tilde x})\partial_{\tilde x}-3\mu^2 \sec^2(\mu {\tilde x}){\tilde v}\partial_{\tilde v},\\[1ex]
c>0,\,d=0\colon\quad&
\partial_{\tilde t}-3\mu\tanh(\mu {\tilde x})\partial_{\tilde x}+3\mu^2 {\rm sech}^2(\mu {\tilde x}){\tilde v}\partial_{\tilde v},\\[1ex]
&\partial_{\tilde t}-3\mu\coth(\mu {\tilde x})\partial_{\tilde x}-3\mu^2 {\rm cosech}^2(\mu {\tilde x}){\tilde v}\partial_{\tilde v}.
\end{array}
\]

Using PET~\eqref{EqHodographXV2}, we are able to obtain
potential nonclassical symmetries of equation~\eqref{eqPorMedEq} with
$n=-1$, $m=1$ and $f=ax^3+bx^2+cx+d$, where $a\ne0$ and we assume $b=0$,
which are associated with potential system~\eqref{sysPotSys1PorMedEq}:
\[\arraycolsep=0ex
\begin{array}{ll}
\lefteqn{a<0\colon}\phantom{c=0,\,d=0\colon\quad}%
\partial_ t-\frac32(ax^3+cx+d)\partial_x\pm\frac32\sqrt{-2a}\,x\partial_v+\frac32[3ax^2u+cu\pm\sqrt{-2a}]\partial_u,\\[1ex]
c=0,\,d=0\colon\quad\partial_t+3v^{-2}x\partial_x+3v^{-1}\partial_v+6uv^{-3}(xu-v)\partial_u,\\[1ex]
c<0,\,d=0\colon\\[1ex]
\partial_t+3\mu^2 \sec^2(\mu v)x\partial_x-3\mu\tan(\mu v)\partial_v-6\mu^2 \sec^2(\mu v)[\mu xu\tan(\mu v)+1]u\partial_u,\\[1ex]
c>0,\,d=0\colon\\[1ex]
\partial_t-3\mu^2 {\rm sech}^2(\mu v)x\partial_x+3\mu\tanh(\mu v)\partial_v-6\mu^2 {\rm sech}^2(\mu v)[\mu xu\tanh(\mu v)-1]u\partial_u,
\\[1ex]
\partial_t+3\mu^2 {\rm cosech}^2(\mu v)x\partial_x+3\mu\coth(\mu v)\partial_v+6\mu^2 {\rm cosech}^2(\mu v)[\mu xu\coth(\mu v)-1]u\partial_u.
\end{array}
\]
Note that the operator presented for the case $a<0$ is in fact usual nonclassical symmetry since it is projectable to the space $(t,x,u)$.

There exist two ways to use mappings between equations in the investigation of nonclassical symmetries.
Suppose that nonclassical symmetries of the imaged equation are known.
The first way is to take the preimages of the constructed operators.
Then we can reduce the initial equation with respect to the preimaged operators
to find its non-Lie exact solutions.
The above way seems to be non-optimal since the ultimate goal of the investigation of nonclassical symmetries is
the construction of exact solutions.
This observation is confirmed by the fact that the imaged equation
and the associated nonclassical symmetry operators
often have a simpler form and therefore, are more suitable than their preimages.
This is why the second way based on the implementation of reductions in the imaged equation and preimaging the
obtained exact solutions instead of preimaging the corresponding reduction operators is preferable.
The same observation is true for Lie symmetries.

For example, the equation $u_t=[(u^{-1})_{x}-x^3u]_x$ is mapped by~\eqref{EqHodographXV2}
to the equation $\tilde v_{\tilde t}={\tilde v}_{\tilde x\tilde x}-\tilde v^3$.
Therefore, we have the mapping between solutions
\[
\tilde v=\dfrac{2\sqrt{2}\, \tilde x}{{\tilde x}^2+6\tilde t}\quad\rightarrow\quad
u=-\sqrt{2}\dfrac{\sqrt{1+3tx^2}\pm1}{x^2\sqrt{1+3tx^2}}.
\]
Analogously, the equation $u_t=[(u^{-1})_{x}-x(x^2-1)u]_x$ is transformed
via~\eqref{EqHodographXV2} to the equation
$\tilde v_{\tilde t}={\tilde v}_{\tilde x\tilde x}-\tilde v{}^3+\tilde v$.
Under the inverse transformation, the known non-Lie exact solution~\cite{ArrigoHillBroadbridge1993,Clarkson&Mansfield1993}
\[\tilde v=\frac{C_1\exp\left(\frac{\sqrt{2}}2\tilde x\right)-C_1'\exp\left(-\frac{\sqrt{2}}2\tilde x\right)}
{C_2\exp\left(-\frac32\tilde t\right)+C_1\exp\left(\frac{\sqrt{2}}2\tilde x\right)
+C_1'\exp\left(-\frac{\sqrt{2}}2\tilde x\right)}
\]
of the imaged equation is mapped to the exact solution
\[
u=\sqrt{2}\ln\left|\dfrac{C_2xe^{\frac32t}\pm\sqrt{C_2^2x^2e^{3t}-C_3(x^2-1)}}{C_1(x-1)}\right|
\]
of the initial equation.

\medskip

{\bf\mathversion{bold}$f=x^{-2}$, $n=-1,$ $m=-2$} (case 1).
The reducing transformation in this case is the most complicated:
\begin{equation*}\hspace*{-.5\arraycolsep}\begin{array}{lll}
v_x=u, \  v_t=-u^{-2}u_x+x^{-2}u^{-2} &\sim&u_t=((u^{-1})_x+x^{-2}u^{-2})_x
\\ \displaystyle
\Bigg\downarrow\ \lefteqn{\tilde t=-t,\  \tilde x=v+\frac t3,\ \tilde u=\frac1{xu}+\frac13,\  \tilde v=\ln x+\frac v3 +\frac2{27}t}&& \\
\tilde v_{\tilde x}=\tilde u,\ \tilde v_{\tilde t}={\tilde u}_{\tilde x}+\tilde u^3 &\sim&
\tilde v_{\tilde t}=\tilde v_{\tilde x\tilde x}+\tilde v_{\tilde x}^3
\end{array}\end{equation*}
The imaged `potential' equation $\tilde v_{\tilde t}=\tilde v_{\tilde x\tilde x}+\tilde v_{\tilde x}^3$
possesses only trivial shift and scale Lie symmetries.
Namely, its maximal Lie invariance algebra is generated by the operators $\p_t$, $\p_x$, $\p_v$ and $4t\p_t+2x\p_x+v\p_v$.
Its first prolongation with respect to~$x$ gives the maximal Lie invariance algebra
$\langle\p_t,\ \p_x,\ \p_v,\ 4t\p_t+2x\p_x+v\p_v-u\p_u\rangle$ of the imaged potential system.
Usually such simple operators do not induce purely potential symmetries via potential equivalence transformations
similar to the potential hodograph transformation.
This is not the case here since due to the complexity of the applied transformation
the preimage of the scale operator $4t\p_t+2x\p_x+v\p_v-u\p_u$ is a purely potential symmetry.

\section{Conclusion}

The complete classification of potential symmetries for a given partial differential equation
can be achieved by considering all the potential systems corresponding to the finite-dimensional subspaces
of the space of conservation laws of the equation.
Further study can be introduced in determining conservation laws for the potential systems which, in turn,
lead to the second generation of potential systems.
Lie symmetries can be derived for the  second generation of potential
systems with the optimal goal to obtain potential symmetries for the original equation.
This procedure can be iterated.

The so-called hidden potential symmetries that appear recently
in \cite{Gandarias2008} are ordinary potential symmetries in the primary sense of~\cite{Bluman&Kumei1989,Bluman&Reid&Kumei1988}.
They can be obtained using the standard approach described in the present paper
and have been earlier found for a number of different equations.
Moreover, these symmetries are simplest ones since each of them involves a single potential.
The attribute ``hidden'' serves in~\cite{Gandarias2008}
for emphasizing a difference between the cases of constant and nonconstant characteristics
(``natural'' and ``hidden'' potential systems, respectively)
but this difference is not essential in any way.
We point out that, to our knowledge, hidden potential symmetries did not appear in the literature before.
As a rule, the attribute ``hidden'' is used for symmetries of a system of differential equations,
which in fact are symmetries of a related system in certain ``usual'' sense.
Thus, hidden point symmetries of a system of partial differential equations arise as point symmetries
of systems obtained by Lie reductions of the initial one~\cite[p.~197]{Olver1993}.
For details the reader can refer to~\cite{Abraham-Shrauner&Govinder2006}.
The first nontrivial example of such hidden symmetries was found by Kapitansky~\cite{Kapitanskiy1978} for the Euler equations.
It is also presented in~\cite{Olver1993}.
Wide families of such hidden symmetries of the Navier--Stokes and Euler equations were constructed in
\cite{Fushchych&Popovych1994,PopovychH2004}.
Different notions of hidden symmetries were invented for ordinary differential
equations~\cite{Abraham-Shrauner&Leach&Govinder&Ratcliff1995}.
They are connected with lowering or increasing the order of the corresponding equations via differential substitutions.
In~\cite{Fushchich&Nikitin1987} another notion of hidden symmetries was proposed for linear partial differential equations
with extending the class of admissible symmetry operators by pseudo-differential ones.

Hidden symmetries of all the above kinds are stable under point transformations in the sense that
the image of a hidden symmetry of an initial object (a differential equation or a system of such equations)
under a point transformation is a hidden symmetry of the corresponding transformed object.
This property is important since involving different point equivalence relations in statements and solutions of problems
is one of the main principles of group analysis of differential equations.
Since the ``hidden'' potential symmetries that appear in \cite{Gandarias2008} do not possess
the property of the stability with respect to point transformations,
the usage of the attribute ``hidden'' is not justified therein.

Another point which should be emphasized is that point and potential equivalence transformations
in classes of differential equations play an important role in the investigation and application of
potential frames over them.
Often new conservation laws, potential symmetries and corresponding invariant solutions can be
easier constructed from known ones via such transformations than by direct calculations.
In the present paper this conclusion has been illustrated by a number of examples.

\subsection*{Acknowledgements}

The research of NMI and OOV was partially supported by the Grant of the President of Ukraine for young scientists
(project number GP/F26/0005).
The research of ROP was supported  by the Austrian Science Fund (FWF), START-project Y237
and project P20632.
NMI, ROP and OOV are grateful for the hospitality shown during their visits to the University of Cyprus.
The authors thank all four referees for a lot of helpful remarks and suggestions.


\begin{thebibliography}{99}\itemsep=-.1ex
\footnotesize

\bibitem{Abraham-Shrauner&Govinder2006}
Abraham-Shrauner B., Govinder K.S.,
Provenance of type II hidden symmetries from nonlinear partial differential equations,
{\it J. Nonlinear Math. Phys.}, 2006, V.13,  612--622.

\bibitem{Abraham-Shrauner&Leach&Govinder&Ratcliff1995}
Abraham-Shrauner B., Leach P.G.L., Govinder K.S. and Ratcliff G.,
Hidden and contact symmetries of ordinary differential equations,
{\it J. Phys. A}, 1995, V.28, 6707--6716.

\bibitem{Akhatov&Gazizov&Ibragimov1989}
Akhatov~I.Sh., Gazizov~R.K. and Ibragimov~N.Kh.,
Nonlocal symmetries. A heuristic approach,
{\it Itogi Nauki i Tekhniki, Current problems in mathematics. Newest results}, 1989, V.34, 3--83
(Russian, translated in {\it J. Soviet Math.}, 1991, V.55, 1401--1450).

\bibitem{Anco&Bluman2002a}
Anco~S.C. and Bluman~G.,
Direct construction method for conservation laws of partial differential equations.~I. Examples of conservation law classifications,
{\it Eur. J. Appl. Math.}, 2002, V.13, Part 5, 545--566; arXiv:math-ph/0108023.

\bibitem{Anco&Bluman2002b}
Anco~S.C. and Bluman~G.,
Direct construction method for conservation laws of partial differential equations.~II. General treatment,
{\it Eur. J. Appl. Math.}, 2002, V.13, Part 5, 567--585; arXiv:math-ph/0108024.

\bibitem{ArrigoHillBroadbridge1993}
Arrigo D.J.,  Hill J.M., Broadbridge P.,
Nonclassical symmetry reductions of the linear diffusion equation with a nonlinear source,
{\it IMA J. Appl. Math.}, 1994, V.52,  1--24.

\bibitem{Bluman&Kumei1989}
Bluman~G.W., Kumei~S., Symmetries and Differential Equations, Springer, New York, 1989.

\bibitem{Bluman&Reid&Kumei1988}
Bluman~G.W., Reid~G.J. and Kumei~S.,
New classes of symmetries for partial differential equations,
{\it J. Math. Phys.}, 1988, V.29, 806--811.

\bibitem{bluman93a}
Bluman G.W.,
Use and construction of potential symmetries,
{\it Math. Comput. Modelling}, 1993, V.18, 1--14.

\bibitem{bluman93b}
Bluman G.W., Potential symmetries and equivalent conservation laws,
in: N.H. Ibragimov, M. Torrisi, A. Valenti (Eds.), Modern Group Analysis:
Advanced Analytical and Computational Methods in Mathematical Physics (Acireale, 1992), Kluwer, Dordrecht, 1993, pp. 71–-84.

\bibitem{Bryant&Griffiths1995b}
Bryant~R.L. and Griffiths~P.A.,
Characteristic cohomology of differential systems II: Conservation laws for a class of parabolic equations,
{\it Duke Math. J.}, 1995, V.78, 531--676.

\bibitem{Clarkson&Mansfield1993}
Clarkson P.A., Mansfield E.L.,
Symmetry reductions and exact solutions of a class of nonlinear heat equations,
{\it Physica D}, 1994, V.70,  250--288.

\bibitem{Dorodnitsyn1979}
Dorodnitsyn V.A., Group properties and invariant solutions of a
nonlinear heat equation with a source or a sink, Preprint N~57,
Moscow, Keldysh Institute of Applied Mathematics of Academy of
Sciences USSR, 1979.

\bibitem{Fushchich&Nikitin1987}
Fushchich W.I. and Nikitin A.G.,
{\it Symmetries of Maxwell's equations}, D.Reidel, Dordrecht, 1987.

\bibitem{Fushchych&Popovych1994}
Fushchych W.I. and Popovych R.O.,
Symmetry reduction and exact solution of the Navier-Stokes equations,
{\it J. Nonlinear Math. Phys.}, 1994, V.1, 75--113, 156--188; arXiv:math-ph/0207016.

\bibitem{FushchichSerov1990}
Fushchich W.I., Serov N.I., Conditional invariance and reduction of nonlinear heat equation,
{\it Dokl. Akad. Nauk Ukrain. SSR Ser. A}, 1990, no.7, 24--27 (in Russian).
Availiable at http://www.imath.kiev.ua/$\sim$fushchych.

\bibitem{Gandarias1996}
Gandarias M.L.,
Potential symmetries of a porous medium equation,
{\it J. Phys. A: Math. Gen.}, 1996, V.29, 5919--5934.

\bibitem{Gandarias2008}
Gandarias M.L.,
New potential symmetries for some evolution equations,
{\it Physica A}, 2008,  V.387, 2234--2242.

\bibitem{Ibragimov1994V1}
Ibragimov N.H. (Editor), Lie group analysis of differential
equations --- symmetries, exact solutions and conservation laws,
V.1, CRC Press, Boca Raton, FL, 1994.

\bibitem{Ivanova&Popovych2007CommentOnMei}
Ivanova N.M. and Popovych R.O.,
Equivalence of conservation laws and equivalence of potential systems,
{\it Internat. J. Theor. Phys}, 2007, V.46, 2658--2668;
preprint N1885 of ESI for Mathematical Physics, \mbox{arXiv:math-ph/0611032}.

\bibitem{Ivanova&Popovych&Sophocleous2004}
Ivanova~N.M., Popovych~R.O. and Sophocleous~C.,
Conservation laws of variable coefficient diffusion--convection equations,
{\it Proceedings of Tenth International Conference in Modern Group Analysis}, (Larnaca, Cyprus, 2004), 107--113.

\bibitem{Ivanova&Popovych&Sophocleous2006Part1}
Ivanova~N.M., Popovych~R.O. and Sophocleous~C.,
Group analysis of variable coefficient diffusion--convection equations. I. Enhanced group classification, 
2007, arXiv:0710.2731.

\bibitem{Ivanova&Popovych&Sophocleous2006Part3}
Ivanova~N.M., Popovych~R.O. and Sophocleous~C.,
Group analysis of variable coefficient diffusion--convection equations. III. Conservation laws, 
2007, arXiv:0710.3053.

\bibitem{Ivanova&Popovych&Sophocleous2006Part4}
Ivanova~N.M., Popovych~R.O. and Sophocleous~C.,
Group analysis of variable coefficient diffusion--convection equations. IV. Potential symmetries,
2007, arXiv:0710.4251.

\bibitem{Kapitanskiy1978}
Kapitanskiy L.V.,
Group analysis of the Navier--Stokes and Euler equations in the presence
of rotational symmetry and new exact solutions of these equations,
{\it Dokl. Acad. Sci. USSR}, 1978, V.243, 901--904.

\bibitem{Lie1881}
Lie S., \"Uber die Integration durch bestimmte Integrale von einer Klasse linear partieller
Differentialgleichung, {\it Arch. for Math.}, 1881, V.6, N~3, 328--368.
(Translation by N.H. Ibragimov:
Lie S. On integration of a class of linear partial differential equations by means of
definite integrals, {\it CRC Handbook of Lie Group Analysis of Differential Equations},
Vol. 2, 1994, 473--508).

\bibitem{Meleshko1994}
Meleshko  S.V.,
Group classification of equations of two-dimensional gas motions,
{\it Prikl. Mat. Mekh.}, 1994, V.58, 56--62 (in Russian);
translation in {\it J.~Appl. Math. Mech.}, 1994, V.58, 629--635.

\bibitem{Mikhailov&Shabat&Sokolov1991}
Mikhailov A.V., Shabat A.B. and Sokolov V.V.,
The symmetry approach to classification of integrable equations (pp. 115--184), in {\it What is integrability?},
Edited by V.E. Zakharov. Springer Series in Nonlinear Dynamics. Springer-Verlag, Berlin, 1991.

\bibitem{Olver1993}
Olver P.J., Applications of Lie groups to differential equations.
Second edition. Graduate Texts in Mathematics, 107. Springer-Verlag, New York, 1993.

\bibitem{Ovsiannikov1982}
Ovsiannikov~L.V., Group analysis of differential equations,
Academic Press, New York, 1982.

\bibitem{PopovychH2004}
Popovych H.V.,
Lie, partially invariant, and nonclassical submodels of Euler equations,
{\it Proceedings of Institute of Mathematics of NAS of Ukraine}, 2002, V.43, Part 1, 178--183.

\bibitem{Popovych2008a}
Popovych R.O., Reduction operators of linear second-order parabolic equations,
{\it J. Phys. A}, 2008, V.41, 185202; arXiv:0712.2764.

\bibitem{Popovych&Ivanova2004ConsLawsLanl}
Popovych R.O. and Ivanova N.M.,
Hierarchy of conservation laws of diffusion--convection equations,
{\it J. Math. Phys.}, 2005, V.46, 043502, arXiv:math-ph/0407008.

\bibitem{Popovych&Ivanova2005PETs}
Popovych R.O. and Ivanova N.M., Potential equivalence transformations for nonlinear diffusion--convection equations,
{\it J. Phys. A},  2005, V.38, 3145--3155, arXiv:math-ph/0402066.

\bibitem{Popovych&Kunzinger&Eshraghi2006}
Popovych R.O., Kunzinger M., Eshraghi H.,
Admissible transformations and normalized classes of nonlinear Schr\"odinger equations, {\it Acta Appl. Math.}, doi:10.1007/s10440-008-9321-4, 45 p.,
arXiv:math-ph/0611061.

\bibitem{Popovych&Kunzinger&Ivanova2008}
Popovych R.O., Kunzinger M. and Ivanova N.M.,
Conservation laws and potential symmetries of linear parabolic equations,
{\it Acta Appl. Math.}, 2008, V.100, no.2, 113--185, arXiv:0706.0443.

\bibitem{Pucci&Saccomandi1993}
Pucci E. and Saccomandi G.,
Potential symmetries and solutions by reduction of partial differential equations
{\it J. Phys. A}, 1993, V.26 681--690.

\bibitem{Pucci&Saccomandi1993b}
Pucci E. and Saccomandi G., Potential symmetries of Fokker Plank equations,
Modern Group Analysis: Advanced Analytical and Computational Methods in Mathematical Physics (Acireale, 1992),
Kluwer Acad. Publ., Dordrecht, 1993, 291--298.

\bibitem{Saccomandi1997}
Saccomandi G.,
Potential symmetries and direct reduction methods of order two,
{\it J. Phys. A}, 1997, V.30, 2211--2217.

\bibitem{Sophocleous2000}
Sophocleous~C.,
Potential symmetries of inhomogeneous nonlinear diffusion equations,
{\it Bull. Austral. Math. Soc.}, 2000, V.61, 507--521.

\bibitem{Sophocleous2003}
Sophocleous~C.,
Classification of potential symmetries of generalised inhomogeneous nonlinear diffusion equations,
{\it Physica A}, 2003, V.320, 169--183.

\bibitem{Sophocleous2005}
Sophocleous~C.,
Further transformation properties of generalised inhomogeneous nonlinear diffusion equations with variable coefficients,
{\it Physica A}, 2005, V.345, 457--471.

\bibitem{VJPS2007}
Vaneeva~O.O., Johnpillai A.G., Popovych~R.O. and Sophocleous~C.,
Enhanced group analysis and conservation laws of variable
coefficient reaction--diffusion equations with power nonlinearities,
{\it J. Math. Anal. Appl.}, 2007, V.330, 1363--1386; arXiv:math-ph/0605081.

\bibitem{VPS2008}
Vaneeva~O.O., Popovych~R.O. and Sophocleous~C.,
Enhanced group analysis and exact solutions  of variable coefficient semilinear
diffusion equations with a power source,
{\it Acta Appl. Math.}, doi:10.1007/s10440-008-9280-9, 46 p., arXiv:0708.3457.

\bibitem{Wolf2002}
Wolf~T.,
A comparison of four approaches to the calculation of conservation laws,
{\it Eur. J. Appl. Math.}, 2002, V.13, Part 5, 129--152.

\bibitem{Zhdanov&Lahno2005}
Zhdanov R. and Lahno V.,
Group Classification of the General Evolution Equation: Local and Quasilocal Symmetries
{\it SIGMA}, 2005, V.1 Paper 009, 7 pages; arXiv:nlin.SI/0510003.

\end{thebibliography}
\end{document}